\documentstyle[psfig,xspace]{mn}

\newif\ifAMStwofonts
\AMStwofontstrue
\def\be{\begin{equation}}
\def\ee{\end{equation}}
\def\etal{{\it et al.~\/}}

\def\ie{{\it i.e.~\/}}
\def\eg{{\it e.g.\/}}

\def\tcmb{T_{CMB}}
\def\tcmb0{T_{CMB}^0}
\def\ltsima{$\; \buildrel < \over \sim \;$}
\def\simlt{\lower.5ex\hbox{\ltsima}}
\def\gtsima{$\; \buildrel > \over \sim \;$}
\def\simgt{\lower.5ex\hbox{\gtsima}}

\def\h2{H$_2$\xspace} 
\def\m{$^{-1}$\xspace}
\def\mm{$^{-2}$\xspace}
\def\mmm{$^{-3}$\xspace}

\ifoldfss

  \ifCUPmtlplainloaded \else
    \NewTextAlphabet{textbfit} {cmbxti10} {}
    \NewTextAlphabet{textbfss} {cmssbx10} {}
    \NewMathAlphabet{mathbfit} {cmbxti10} {} 
    \NewMathAlphabet{mathbfss} {cmssbx10} {} 
  \fi
  \ifAMStwofonts
    \ifCUPmtlplainloaded \else
      \NewSymbolFont{upmath} {eurm10}
      \NewSymbolFont{AMSa} {msam10}
      \NewMathSymbol{\upi}     {0}{upmath}{19}
      \NewMathSymbol{\umu}     {0}{upmath}{16}
      \NewMathSymbol{\upartial}{0}{upmath}{40}
      \NewMathSymbol{\leqslant}{3}{AMSa}{36}
      \NewMathSymbol{\geqslant}{3}{AMSa}{3E}

       \let\le=\leqslant
       \let\ge=\geqslant
    \fi
  \fi
\fi 

\ifnfssone
  \newmathalphabet{\mathit}
  \addtoversion{normal}{\mathit}{cmr}{m}{it}
  \addtoversion{bold}{\mathit}{cmr}{bx}{it}

  \newmathalphabet{\mathbfit} 
  \addtoversion{normal}{\mathbfit}{cmr}{bx}{it}
  \addtoversion{bold}{\mathbfit}{cmr}{bx}{it}
  \newmathalphabet{\mathbfss} 
  \addtoversion{normal}{\mathbfss}{cmss}{bx}{n}
  \addtoversion{bold}{\mathbfss}{cmss}{bx}{n}
  \ifAMStwofonts
    \ifCUPmtlplainloaded \else
      \UseAMStwoboldmath
      \makeatletter
      \new@mathgroup\upmath@group
      \define@mathgroup\mv@normal\upmath@group{eur}{m}{n}
      \define@mathgroup\mv@bold\upmath@group{eur}{b}{n}
      \edef\UPM{\hexnumber\upmath@group}
      \new@mathgroup\amsa@group
      \define@mathgroup\mv@normal\amsa@group{msa}{m}{n}
      \define@mathgroup\mv@bold\amsa@group{msa}{m}{n}
      \edef\AMSa{\hexnumber\amsa@group}
      \makeatother
      \mathchardef\upi="0\UPM19
      \mathchardef\umu="0\UPM16
      \mathchardef\upartial="0\UPM40
      \mathchardef\leqslant="3\AMSa36
      \mathchardef\geqslant="3\AMSa3E

       \let\le=\leqslant
       \let\ge=\geqslant
    \fi
  \fi
\fi 

\ifnfsstwo

  \newcommand{\placefigure}[1]{}
  \DeclareMathAlphabet{\mathbfit}{OT1}{cmr}{bx}{it}
  \SetMathAlphabet\mathbfit{bold}{OT1}{cmr}{bx}{it}
  \DeclareMathAlphabet{\mathbfss}{OT1}{cmss}{bx}{n}
  \SetMathAlphabet\mathbfss{bold}{OT1}{cmss}{bx}{n}
  \ifAMStwofonts
    \ifCUPmtlplainloaded \else
      \DeclareSymbolFont{UPM}{U}{eur}{m}{n}
      \SetSymbolFont{UPM}{bold}{U}{eur}{b}{n}
      \DeclareSymbolFont{AMSa}{U}{msa}{m}{n}
      \DeclareMathSymbol{\upi}{0}{UPM}{"19}
      \DeclareMathSymbol{\umu}{0}{UPM}{"16}
      \DeclareMathSymbol{\upartial}{0}{UPM}{"40}
      \DeclareMathSymbol{\leqslant}{3}{AMSa}{"36}
      \DeclareMathSymbol{\geqslant}{3}{AMSa}{"3E}

       \let\le=\leqslant
       \let\ge=\geqslant
    \fi
  \fi
\fi 

\ifCUPmtlplainloaded \else
  \ifAMStwofonts \else 
    \def\upi{\pi}
    \def\umu{\mu}
    \def\upartial{\partial}
  \fi
\fi

\title{What regulates the velocity distribution of interstellar clouds?}
\author[M. Ricotti \& A. Ferrara]
{Massimo Ricotti$^1$ and  Andrea Ferrara$^2$\\ 
$^1$ Center for Astrophysics and Space Astronomy\\ University of
Colorado,
Boulder, CO 80309\\
$^2$Osservatorio Astrofisico di Arcetri \\ Largo E. Fermi, 5 - 50125
Firenze, Italy\\ }
\date{}
\pagerange{\pageref{firstpage}--\pageref{lastpage}}
\pubyear{2002}
\begin{document}
\maketitle
\label{firstpage}
\begin{abstract}
  Kinetic energy stored in ISM bulk/turbulent motions is a crucial
  ingredient to properly describe most properties of observed
  galaxies.  By using Monte Carlo simulations, we investigate how this
  energy is injected by supernovae and dissipated via cloud collisions
  and derive the corresponding ISM velocity probability distribution
  function (PDF).  The functional form of the PDF for the modulus of
  the velocity dispersion is
  $$p(v) \propto v^2 \exp[-(v/\sigma)^\beta].$$
  The power-law index of
  the PDF depends only on the value of the average cloud collision
  elasticity $\langle \epsilon \rangle$ as $\beta = 2
  \exp(\langle\epsilon \rangle -1)$.  If $\beta$ and the gas velocity
  dispersion $\sigma$ are known, the specific kinetic energy
  dissipated by collisions is found to be \hbox{$\propto \sigma^2 \ln
    (2 / \beta)/(\beta-0.947)$}; in steady state, this is equal to the
  energy input from SNe.  We predict that in a multiphase, low
  metallicity ($Z \approx 5 \times 10^{-3} Z_\odot$) ISM the PDF
  should be close to a Maxwellian ($\beta = 2$) with velocity
  dispersion $\sigma \simgt 11$ km~s\m; in more metal rich systems ($Z
  \simgt 5 \times 10^{-2} Z_\odot$), instead, we expect to observe
  almost exponential PDFs.  This is in good agreement with a number of
  observations that we review and might explain the different
  star formation modes seen in dwarfs and spiral galaxies.
\end{abstract}

\section{Introduction}

An increasing number of recent studies have come to the conclusion
that a large, if not dominant, fraction of the energy injected into
the interstellar medium (ISM) in various forms by stellar activity,
resides in bulk/turbulent motions of the gas (e.g. McKee 1990, Lockman
\& Gehman 1991, Ferrara 1993, Vazquez-Semadeni \etal 1995, Norman \&
Ferrara 1996 [NF]).  Energy in this form is required, for example, to
reproduce the observed vertical distribution of the HI in the Galactic
disk.  The corresponding turbulent pressure is estimated to be from
$\ge 8$ (McKee 1990) to $\simeq 30$ (NF) times larger than the thermal
one. Note that such large ratios imply a high porosity factor, $Q$, of
the hot gas produced by supernova (SN) explosions: McKee (1990)
estimates that $Q \sim 1.1 (P_{turb} / P_{tot})^{4/3}$.  This
dynamical property of the ISM has a dramatic importance in shaping the
resulting structure of the parent galaxy.  A number of numerical
models (V\'azquez-Semadeni \etal 1995, 1998; Ballesteros-Paredes \etal
1999; Korpi \etal 1999) have been developed in order to describe the
global properties of a turbulent ISM in the Galaxy improving on the
pioneering works by Rosen \etal (1993, 1996) and Rosen \& Bregman
(1995).  Although these studies provide a first approach to the study
of the ISM properties on galactic scales, finite resolution limits
their ability to properly describe shocks arising both as a
consequence of SN explosions and cloud collisions, thus making their
estimates of the kinetic energy dissipation rather uncertain. The same
caveat should be made for the treatment of thermal instabilities (and
hence cloud formation) which requires to spatially resolve the ``Field
length'' (Hennebelle \& Perault 1999).  Galaxy formation studies in a
cosmological context (Katz 1992, Mihos \& Hernquist 1994, Navarro \&
Steinmetz 2000, Springel 2000), have by now firmly established that
star formation cannot be properly regulated by forcing a phase
transition of the gas into the hot phase by thermal energy input from
SN explosions alone. Current computational power limits the spatial
and the time scale resolution of the simulations, implying a too
highly dissipative treatment. Moreover small-scale physical processes
usually cannot be modeled with sufficient detail. A surprisingly
better description of galactic disk properties is obtained once an
heuristic equation of state for the ISM is assumed with parameters
adjusted to simulate the effects of a turbulent pressure. In these
models a fraction of the available energy is invested in modifying the
kinetic energy of the gas (Navarro \& Steinmetz 2000). The recipe for
this approach is relatively simple and it constitutes a promising path
to understand galaxy formation, but it needs to be calibrated against
well known situations as for example the Galactic ISM, where a large
number of high quality observations on the HI dynamics are available.
Finally, Ferrara \& Tolstoy (2000) point out that if energy injection
is mainly regulated by massive stars, then the observed metallicity
range of galaxies must be consistent with what inferred from the ISM
kinetic energy budget.  Along these lines they find that a tight
relation between gas velocity dispersion and metallicity in dwarfs
must exist.

In view of the above situation, we have decided to explore how kinetic
energy is injected by SN explosions and radiated away as converging
gas flows collide at supersonic velocities.  The present work is
mainly building on previous investigations by our group which have
concentrated on the analysis of high-res hydro- and MHD simulations of
cloud collisions (Ricotti, Ferrara \& Miniati 1997 [RFM], Miniati
\etal 1997, 1999). The main effort there was devoted to quantify the
amount of kinetic energy dissipation as a function of the collision
parameters and cloud properties. Here we take a more global, although
simplified, approach to understand the interplay between the two above
mentioned processes (SN explosion and cloud-cloud collisions)
regulating the dynamics/energetics of the turbulent ISM.  To this aim,
we will adopt a statistical description based on a Monte Carlo
approach that allows us to derive a probability distribution function
(PDF) for the ISM velocity dispersions. The PDF is then shown to hold
directly measurable and easily interpreted information about the
kinetic energy regulation in the ISM.

The paper is organized in the following manner. In \S~\ref{sec:model}
we discuss our statistical approach to the modeling of a turbulent ISM
and the model assumptions. In \S~\ref{sec:res} we present the results.
In \S~\ref{sec:obs} we show that our results are in agreement and
can explain a number of observational results, that we review, from
published literature. In \S~\ref{sec:disc} we justify
the simple assumptions of our model and in \S~\ref{sec:sum} we summarize
the results.

\section{Model assumptions}\label{sec:model}

We consider a galaxy of total HI mass $M_{CNM}$ distributed in clouds
of mass $M_c$ and cloud filling factor $f_c$.  The clouds constituting
the Galactic Cold Neutral Medium (CNM) phase have typically a gas
number density $n_c= \rho_c/\mu m_p \sim 50$ cm$^{-3}$, where $\mu$ is
the mean molecular weight and $m_p$ the proton mass. If the clouds are
further idealized as homogeneous spheres of radius $R_c$, the total
number of clouds is $N_{tot}=M_{CNM}/M_c$.  The cross section for
cloud collisions is $A = {\cal B} \pi R_c^2$, where the unknown
parameter ${\cal B} \sim 1$ takes into account the effect of magnetic
field, tidal forces and gravitational focusing, effective at low (5-6
km s\m) cloud relative velocities (Lockman \& Gehman, 1991).  It
follows that the number density of clouds is ${\cal N}_c= 3
N_{tot}f_c/4\pi R_c^3$, and their collisional frequency $\nu={\cal
  N}_c A \overline v_r$, where $\overline v_r$ is the mean relative
velocity of colliding clouds.

We use a set of Monte Carlo simulations to model the
dynamics/dissipation of cloud collisions, and the energy input from SN
explosions and stellar winds.  We assume a homogeneous and isotropic
spatial distribution of the clouds and energy sources. In this case,
the system is completely characterized by the evolution of the
velocity probability distribution function (PDF) for the modulus of
the velocity of the clouds, which in case of a Maxwellian distribution
can be expressed as,
\begin{equation}
p(v) \propto v^2 \exp{\left(-{v^2 \over \sigma^2}\right)}~,
\end{equation}
where the 3D velocity dispersion, $\sigma$, is related to the observed 
1D velocity dispersion by the relationship, $\sigma=\sqrt{3}\sigma_{1D}$.

\begin{figure}
\centerline{\psfig{figure=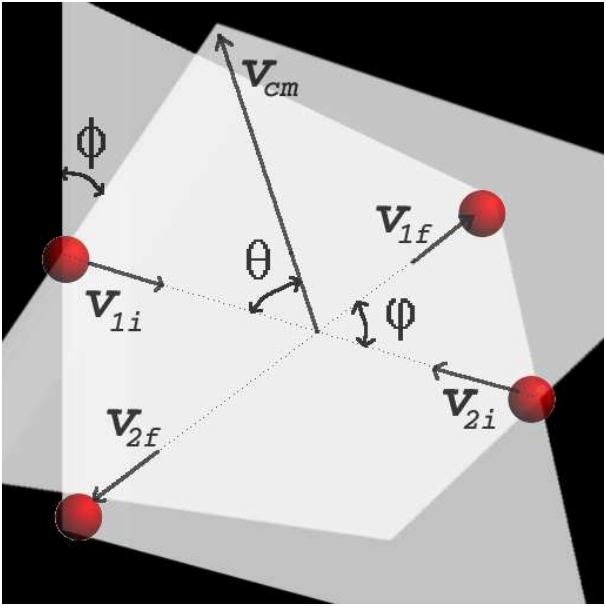,height=8cm}}
\caption{\label{fig:1} Sketch of the collision geometry and angle
definitions.}
\end{figure}

Starting from an initial PDF (typically a Maxwellian), we evolve the
distribution at subsequent time steps $\Delta t=(\nu N_{tot})^{-1}$ as
follows.  At each time step we select two clouds (we do not consider
three body encounters) from the velocity distribution, with initial
velocities $v_{1,i}$ and $v_{2,i}$, by randomly sampling the PDF and
calculate the final velocities $v_{1,f}$ and $v_{2,f}$ after the
collision. The expression for the new velocities are
\begin{eqnarray}
v^2_{1,f}=v_r^2+v_{cm}^2 - 2 v_r v_{cm} \cos{ \varphi} ~;\\
v^2_{2,f}=v_r^2+v_{cm}^2 + 2 v_r v_{cm} \cos{ \varphi} ~,
\end{eqnarray}
where the relative velocity after the collision and the center of mass
velocities are, respectively:
\begin{eqnarray}
v^2_r={\epsilon \over 2}[v_{1,i}^2+v_{2,i}^2- \nonumber\\ 
2 v_{1,i} v_{2,i} (\sin{\theta}\cos{\phi}+\cos{\theta}\sin{\phi}\cos{\varphi})]~;\\
v^2_{cm}={1 \over 2}[v_{1,i}^2+v_{2,i}^2+ \nonumber \\ 
2 v_{1,i} v_{2,i} (\sin{\theta}\cos{\phi}+\cos{\theta}\sin{\phi}\cos{\varphi})]~;
\end{eqnarray}
$\epsilon(v_{r,i},R_c ,Z)$ is the elasticity of the collision, with
$v_{r,i}$ being the relative velocity before the collision (see
discussion in \S~\ref{ssec:ccrev}) and $Z$ the gas metallicity.
For the definition of the angles $\theta, \varphi, \phi$ see the
sketch in Fig.~\ref{fig:1}.

Every $\gamma=(\nu_{SN}\Delta t)^{-1}$ collisions, where $\nu_{SN}$ is
the SN rate in the galaxy, an energy $E=\kappa E_{SN}$ is added to the
system; $E_{SN}=10^{51}$~erg is the total energy of the SN explosion
and $\kappa = 0.02$ is the fraction\footnote{The adopted value of the
  coefficient $\kappa$ is perhaps controversial because based on 1D
  hydro-simulations. In a more realistic model dynamical instabilities
  and turbulence could play a crucial role. We do not worry about this
  uncertainty because the qualitative results of this paper are
  unaffected by a different choice of $\kappa$.}  converted in kinetic
energy of the engulfed clouds (Thornton \etal 1998). As long as cloud
coagulation, cloud destructions or formations are neglected in the
model, the following simple argument suggests that the cloud collision
rate, $\nu$, is constant.  The scale height of the gaseous disk in
galaxies is $H \propto \overline v_r$ (McKee 1990, Ferrara 1993).
Since in our model we assume constant gas mass and cloud mass (\ie,
$N_{tot}=const.$, at least in a statistical sense) we have ${\cal N}_c
\propto \overline v_r^{-1}$.  It follows from the definition of $\nu$,
given at the beginning of this paragraph, that $\nu$ is constant. If,
in addition, we assume constant SN explosion rate we have
$\gamma=const.$ since $\Delta t=(\nu N_{tot})^{-1}$ and $N_{tot}$ is
constant. To summarize, the model is characterized by four constant
parameters which must be appropriately chosen to describe the galaxy
under study: $M_c$, $R_c$, $Z$ and $\gamma$. Note that in a
self-consistent global model of the ISM these four parameters are not
independent but correlated. Here we are not interested in a global
modeling of the ISM that would require a different approach to the
problem using high resolution hydro- simulations. Therefore we rely on
observational and theoretical results found in literature to determine
a range of realistic values for those parameters.
  
\subsection{Energy sources and sinks}   

\subsubsection{Turbulent pumping by supernova explosions}

The main source of the kinetic energy required to support the observed
HI bulk motions in the Galaxy is provided by SN explosions and stellar
winds (McKee 1990, NF). The latter authors calculated the source
function for interstellar turbulence by using a detailed formalism to
account for the interaction between SN shock waves and the clouds,
which are thereby accelerated and put into motion. For brevity, we
only repeat here the main results relevant to this study, deferring
the interested reader to NF for further details.

To simplify our treatment we only consider primary shocks, \ie shocks
associated with the SN explosion itself, as a source of kinetic energy
for the clouds. Therefore we neglect secondary shocks, \ie primary shocks
reflected off the clouds; as seen from Fig. 1 of NF  , their contribution 
is less than 1\% both for SNe and winds.

The distribution function ${P_{i}}^{(1)}(\mu)$ for primary shocks due
to the $i$-th source, where $i=1$ ($i=2$) refers to SNe (winds), can
be written as 

\be {P_{i}}({\cal M}) = {4 \pi \over 3} \alpha_i \gamma_i
{R_{0i}^3\over V} {\cal M}^{-(\alpha + 1)} ={\alpha_i P_{0i} \over
  {\cal M}^{ (\alpha + 1)}}, 
\ee 

where ${\cal M}$ is the Mach number of the shock with respect to the
intercloud gas supposed here to be at $T=8000$~K and $V$ is the volume
affected by the energy input.  The previous expression contains two
parameters ($R_{0i}$ and $\alpha_i$) which completely define the
self-similar stage of the shock radius evolution for the given source.
Adopting the canonical values of NF, we take $R_{0i}/{\rm pc}=(70,70)$
and $\alpha_i=(9/7,9/2)$. The values for the radii of different
turbulence sources (\ie, isolated and clustered SNe, Winds, HII
regions) are given in Table 1 of NF. In turn, the specific value of 70
pc is taken from Cioffi, McKee \& Bertschinger (1988).  The resultant
primary shock distribution in the ISM is then

\be
P({\cal M}) = \sum_i {P_{0i}  \over \mu^{\alpha_i + 1}} 
\ee

From this distribution we randomly extract a value of the shock Mach
number (or velocity, $v_s$); the velocity of the engulfed cloud before
interaction with the shock, $v_{c0}$, is extracted from the actual PDF.
The velocity acquired by the cloud in the direction of the shock
propagation is (McKee \& Cowie 1975, Miesch \& Zweibel 1994)

\be
v_{cs}=\left({\Delta \over \Delta+1}\right){v_{sh} \over \chi^{1/2}}~,\\
\ee
where $\Delta = 4$ for an adiabatic shock and $\chi=100$ is the
canonical value for the density contrast between cloud and intercloud
medium\footnote{$\chi=100$ is an average value for the density
contrast which is essentially derived by imposing pressure equilibrium
between clouds and intercloud medium, as in the classical two-phase
model scheme of the ISM.  For small clouds this ratio can vary;
however, the cloud velocity has only a mild, $\chi ^{-1/2}$,
dependence on its value.}.  Thus the final velocity of the cloud after
the interaction with the shock is
\be
v^2_{ps}=(v_{c0}^2+v_{cs}^2 + 2 v_{cs} v_{c0}\cos \zeta)~,
\ee
where $\zeta$ is the angle between the directions of the shock propagation
and the cloud motion before the interaction. 
  
\subsubsection{Kinetic energy dissipation in the collisions}\label{ssec:ccrev}

Since the pioneering work of Spitzer, interstellar cloud collisions
have always been considered as perfectly inelastic because, for
typical Galactic clouds, the cooling time of the shocked gas is
shorter than the characteristic time scale of the collision.
Nevertheless, this assumption might be rather crude for small clouds
and/or primordial galaxies where the gas metallicity is low and
therefore the cooling time is longer. In a previous paper (RFM), we
have investigated the dependence of the collision elasticity,
$\epsilon_{cc}$, defined as the ratio of the final to the initial
kinetic energy of the clouds, on the velocity and mass ratio of the
colliding clouds, metallicity and magnetic field strength.  The
problem has been studied via high-res numerical simulations. In the
case of head-on collisions, RFM derived a handy analytical
relationship that has been shown to correctly approximate the results
of numerical simulations:
\begin{equation}
\eta_{cc} =1-\epsilon_{cc}= {\Sigma_{cool} \over 6 n_c R_c}\propto {v_r^3
  \over n_c R_c \Lambda(v_r,Z)}~; 
\end{equation}
$\Sigma_{cool}$ is the column density of the post-shock radiative region,
and $\Lambda$ [erg cm$^3$ s$^{-1}$] is the gas cooling function.  For
clarity we summarize here the relevant results of RFM: $(i)$ the
kinetic energy dissipation in cloud collisions is minimum (i.e. the
collision elasticity is maximum) for a cloud relative velocity $v_r
\simeq 30$ km~s$^{-1}$; $(ii)$ the above minimum value is proportional
$Z R_c^2$, where $Z$ is the metallicity and $R_c$ is the cloud size:
the larger is $ZR_c^2$ the more dissipative (\ie, inelastic) the
collision will be; $(iii)$ in general, we find that the energy
dissipation decreases when the magnetic field strength and mass ratio
of the clouds are increased and the metallicity is decreased,
respectively.
  
In principle, molecular hydrogen cooling could be important,
especially in a low metallicity gas.  However, H$_2$ in diffuse gas is
very easily photodissociated by stellar UV flux. Therefore we neglect
to a first approximation the effects of such molecule on the
elasticity.

\begin{figure}
\centerline{\psfig{figure=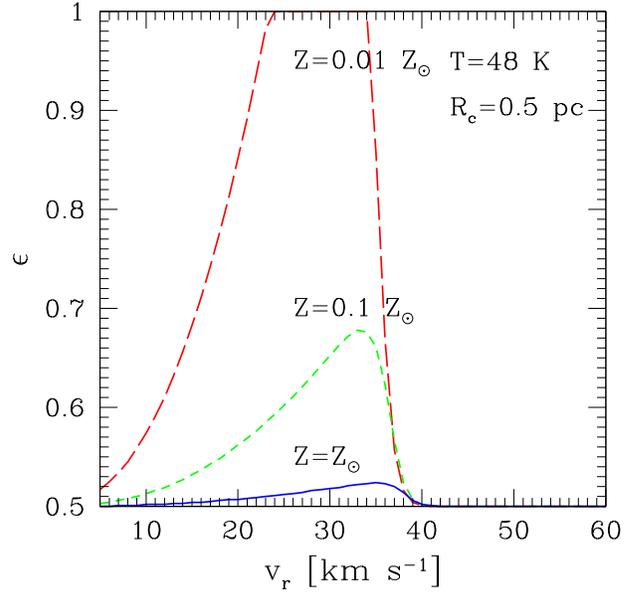,height=9cm}}
\caption{\label{fig:el-v} Elasticity of cloud collisions, $\epsilon$, as
  a function of the cloud relative velocity, $v_{r}$, for clouds with
  metallicities $Z = 1, 0.1$ and 0.01 Z$_\odot$. By definition
  perfectly elastic collisions do not radiate away their initial
  kinetic energy because the shock propagating inside the clouds
  remains always adiabatic. The peak of the cloud collision elasticity
  at $v_r \sim 30$ km s\m corresponds to maximum column density of the
  post-shock radiative region, $\Sigma_{cool} \propto v_r^3/\Lambda$.
  See text for more explanation.}
\end{figure}

In RFM we derived the elasticity for face-on collisions.  We can
estimate the mean elasticity of cloud collisions due to off-center
encounters by assuming that the overlapping cloud area 
experiences a face-on collision whether in the
rest of the cloud volume kinetic energy is conserved. With this
simple hypothesis we obtain 
\begin{equation}
\eta = 1-\epsilon={\eta_{cc} \over 2R_c} \int^{2R_c}_0 \left(1-{b \over
    2R_c}\right)~db={\eta_{cc} \over 2}~;
\label{eq:rfm}
\end{equation}
where $\eta_{cc}$ is the energy dissipated for face-on collisions.
With this definition, perfectly inelastic 3D collisions have a mean
elasticity $\epsilon \simeq 0.5$. In Fig.~\ref{fig:el-v} we show
$\epsilon$ as a function of the cloud relative velocity, $v_{r}$, for
clouds with metallicities $Z = 1, 0.1$ and 0.01 Z$_\odot$. We have
assumed cloud temperature $T=48 K$, cloud radius $R_c =0.5$ pc and
cloud density given by eq.~\ref{eq:peq}, derived form the assumption
of pressure equilibrium of the multiphase ISM. The effect of magnetic
field (neglected here) is both to increase the cross section of the
collision and the elasticity of the face-on collision.

\subsection{Selection of model parameters}                

As previously mentioned the parameters of our model can be fixed
according to the derived observational constraints.  For the Milky Way
the values we adopt here are derived from data collected in Thronson
\& Shull (1990)\nocite{Thronson:90}.  The cloud gas number density is
$n_c \approx 50$ cm$^{-3}$ and the cloud radius is $R_c=0.1-6$~pc; the
cloud masses are therefore in the range $M_c \sim
0.0045-1000$~M$_\odot$. A typical atomic cloud has $R_c=0.5$~pc and
mass $M_c \sim 0.5 $~M$_\odot$. Clouds with $M_c \simgt
1000$~M$_\odot$ are self-gravitating and we assume that they cannot
remain in the atomic phase\footnote{ Even clouds with $M_c \simgt
1000$~M$_\odot$ are fully molecular only in dense knots and should
remain atomic in the outermost regions. But we neglect this
complication.}. The value of $\gamma$, expressing the number of cloud
collisions between two subsequent SN explosions is in the range
$1-1000$. The elasticity of the collision depends on the gas
metallicity $Z$, the cloud radius $R_c$, and relative velocity $v_r$,
according to eq.~\ref{eq:rfm}.  Finally, we have adopted a SN rate of
$\nu_{SN}=0.04$ yr$^{-1}$.  Note that once a value for $\gamma$ is
fixed, the SN rate determines the time step of the simulation but does
not affect the properties of the PDF at equilibrium.

\section{Results}\label{sec:res}

\subsection{A Test Case: Elastic Collisions}

We have first tested our code on the perfectly elastic cloud
collisions case. We have chosen several different initial conditions
for the PDF, and in all cases, after a number of collisions comparable
to the total number of clouds in the box, a stationary Maxwellian PDF
is approached, as expected. The number of clouds and the total energy
are conserved (in this case we set the energy input to zero) within
less than 0.1\%.

\subsection{Constant Elasticity}                

We now assume that the initial PDF is Maxwellian with velocity
dispersion $\sigma_M = 10$ km~s$^{-1}$ for all cases discussed below.
We run several simulations exploring the effects of different values
of $\gamma$, $R_c$, $n_c$ and elasticity. The elasticity range is
$0\le \epsilon \le 1$ and in each simulation is set to a constant
value unrelated to the physical properties of the clouds.

\begin{figure}
\centerline{\psfig{figure=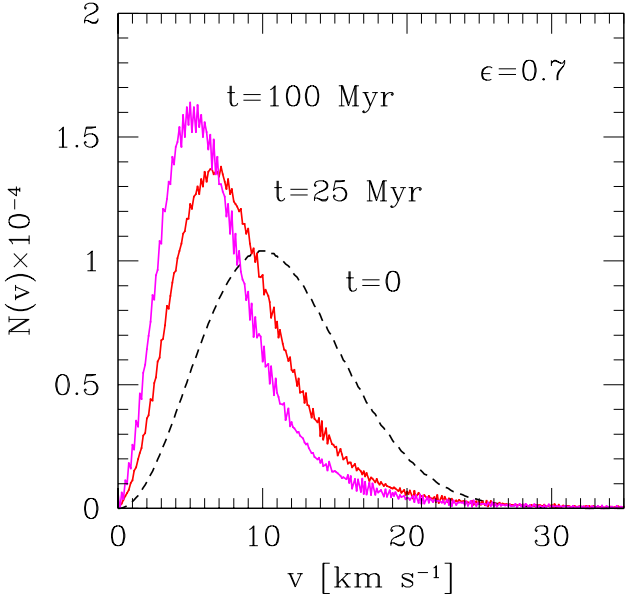,height=9cm}}
\caption{\label{fig:m-c} Evolution of the cloud velocity distribution,
  $N(v)=N_{tot}p(v)$, from a Monte Carlo simulation with
  $\epsilon=0.7, \gamma=110, R_c=4.17~{\rm pc}, M_c=40~{\rm M}_\odot$.
  The dashed line is the initial PDF, a Maxwellian distribution with
  $\sigma_M=10$ km~s$^{-1}$; the solid lines show the PDF at $t=25$ and
  100 Myr. For this run, a stationary PDF is reached after about 50
  Myr.}
\end{figure}

The typical number of collisions required to reach the stationary
regime depends on the mean kinetic energy per cloud mass, $\langle E
\rangle$, at the time when the stationary phase is reached: the lower
is $\langle E\rangle$, the larger is the number of collisions required
to reach the steady state. An example of a Monte Carlo simulation is
shown Fig.~\ref{fig:m-c}.

The shape of the stationary PDF  is well described by the
parametric function
\begin{equation}
p(v)={\beta \over \Gamma(3/\beta) \sigma} \left({v \over
  \sigma}\right)^2 \exp{\left[-\left({v \over
      \sigma}\right)^\beta \right]}~,
\label{eq:gv}
\end{equation}
whose maximum occurs at 
$v_{max}=(2 /\beta)^\beta \,\sigma$. Here $\Gamma(x)$ is the gamma function.
The distribution has total kinetic energy per unit mass
\begin{equation}
\langle E \rangle={\Gamma(5 / \beta) \over
  \Gamma(3 / \beta)}\sigma^2 \sim {3 \over
  2}{1.05 \sigma^2 \over \beta - 0.947}~~~{\rm if}~1<\beta<2~.
\label{eq:sig}
\end{equation}
For the Maxwellian ($\beta=2$), we have $\langle E
\rangle=(3/2)\sigma_M^2$ and $\langle E \rangle=(3/40)\sigma^2$ if
$\beta=1$. In the range $1<\beta<2$ the function $\Gamma(5 /
\beta)/\Gamma(3 / \beta) \approx 1.58/(\beta - 0.947)$.

\begin{figure}
\centerline{\psfig{figure=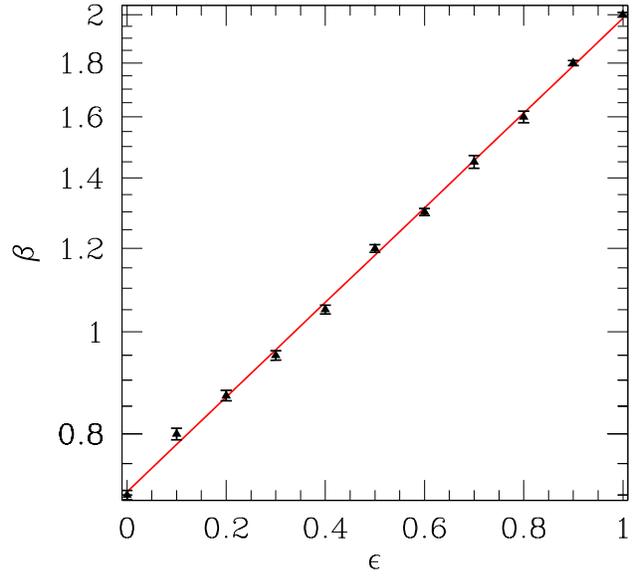,height=9cm}}
\caption{\label{fig:el} Shape of the PDF as a function of the
  elasticity. The parameter $\beta $ in eq.~(\ref{eq:gv}) is a
  function of the elasticity $\epsilon$ only.  The solid line shows the
  curve $\beta = 2 \exp{(\epsilon-1)}$: for elastic collisions
  ($\epsilon= 1$) the PDF is Gaussian, for inelastic collisions
  ($\epsilon = 0.5$) the PDF is close to exponential ($\beta =1.2$).}
\end{figure}

We have performed several simulations to explore the effect of changes
in the model parameters.  If the mass of the clouds is not too large
($\ga 1000$ M$_\odot$), the PDF is given by eq.~(\ref{eq:gv}), with
the index $\beta$ which is a function of the mean elasticity of the
collisions only and independent on the details of the energy
injection. However, in the case of massive clouds each collision
dissipates an energy comparable to the energy injected by a typical SN
explosion. Therefore, at the steady state, the number of cloud
collisions per SN explosion is about one and the functional shape of
the PDF depends on the SN energy input prescription.  In
Fig.~\ref{fig:el} we show the tight correlation of $\beta$ to the
elasticity.  The solid line, showing the best fit to the simulation
points, has the simple expression

\begin{equation}
\beta = 2  e^{(\epsilon-1)}~,~~~~~~0 \le \epsilon \le 1.   
\label{eq:main}
\end{equation}
For the perfectly inelastic case, according to eq.~(\ref{eq:rfm})
$\epsilon=0.5$, implying  $\beta \simeq 1.2$.

A second conclusion can be drawn from our results.  The mean kinetic
energy per unit mass, $\langle E \rangle$, of the equilibrium
distribution is simply related to the energy input per unit mass per cloud
collision, $E_i=\kappa E_{SN}/\gamma M_c$, through the dissipation
parameter $\eta=1-\epsilon$ by the following equation:
\begin{equation}
E_i= \eta \langle E \rangle= (1 - \epsilon) \langle E \rangle =
{\Gamma(5/ \beta)\over \Gamma(3/\beta)}\sigma^2 \ln \left({2 \over 
\beta}\right) 
\label{eq:E}
\end{equation}
where we have used the two above relations.
Eqs.~(\ref{eq:sig})-(\ref{eq:E}) are useful to estimate ISM
parameters, $\langle E \rangle, \epsilon$ and $E_i$ from the observed
shape of the PDF, characterized by $\beta$ and $\sigma$. The typical
value of $M_c$ is a function of metallicity and will be derived in the
next Section.

\placefigure{
\begin{figure}
\centerline{\psfig{figure=pdf_fig5.eps,height=9cm}}
\caption{\label{fig:E} Total energy $\langle E \rangle$ per unit mass 
  of the cloud system as a function of $E_i/\eta$, the ratio of the
  energy input per unit mass and the mean energy dissipated in each
  cloud collision.}
\end{figure}
}

\subsection{Physical Elasticity}

\begin{figure*}
\centerline{\psfig{figure=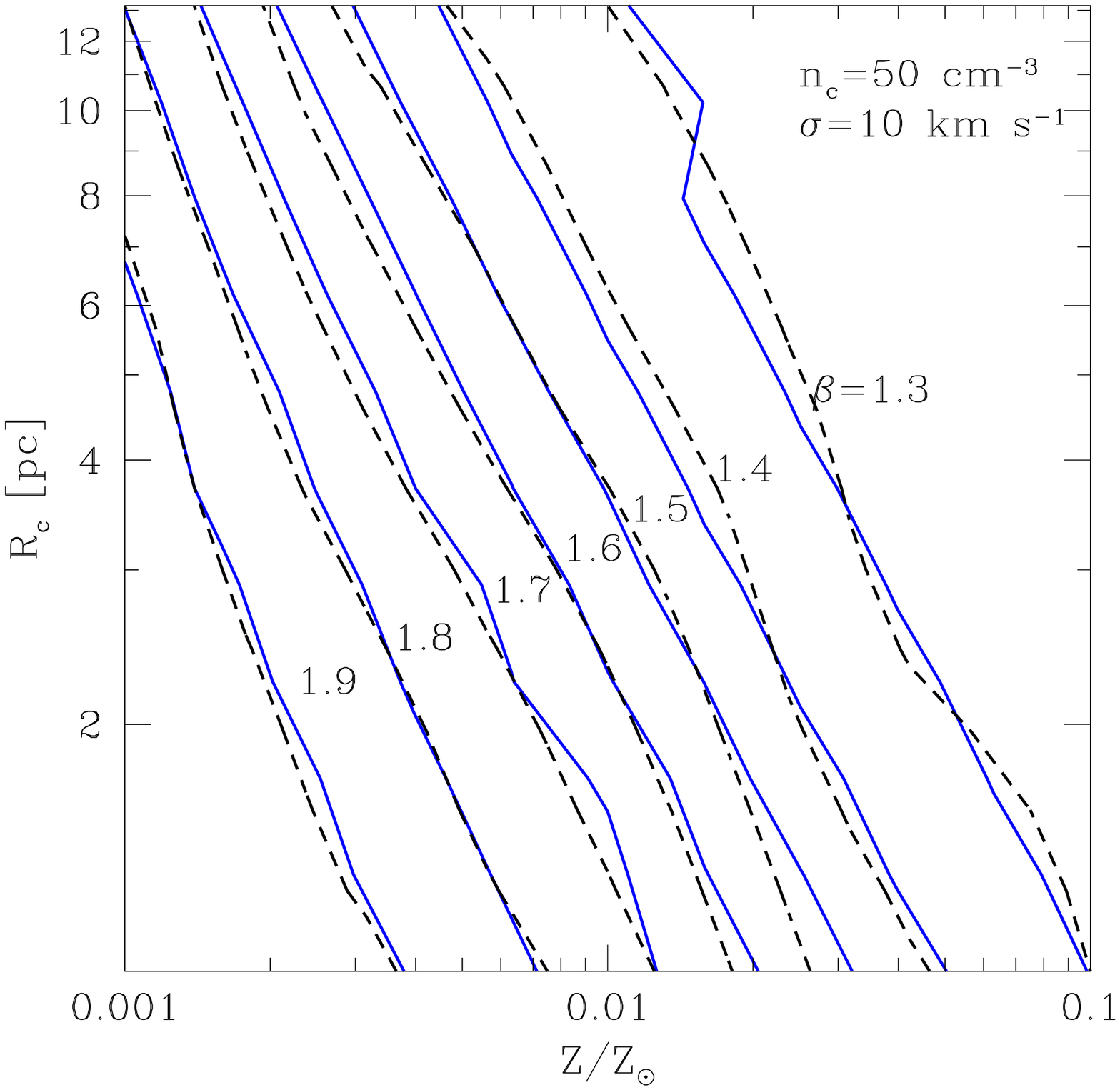,height=8cm} \psfig{figure=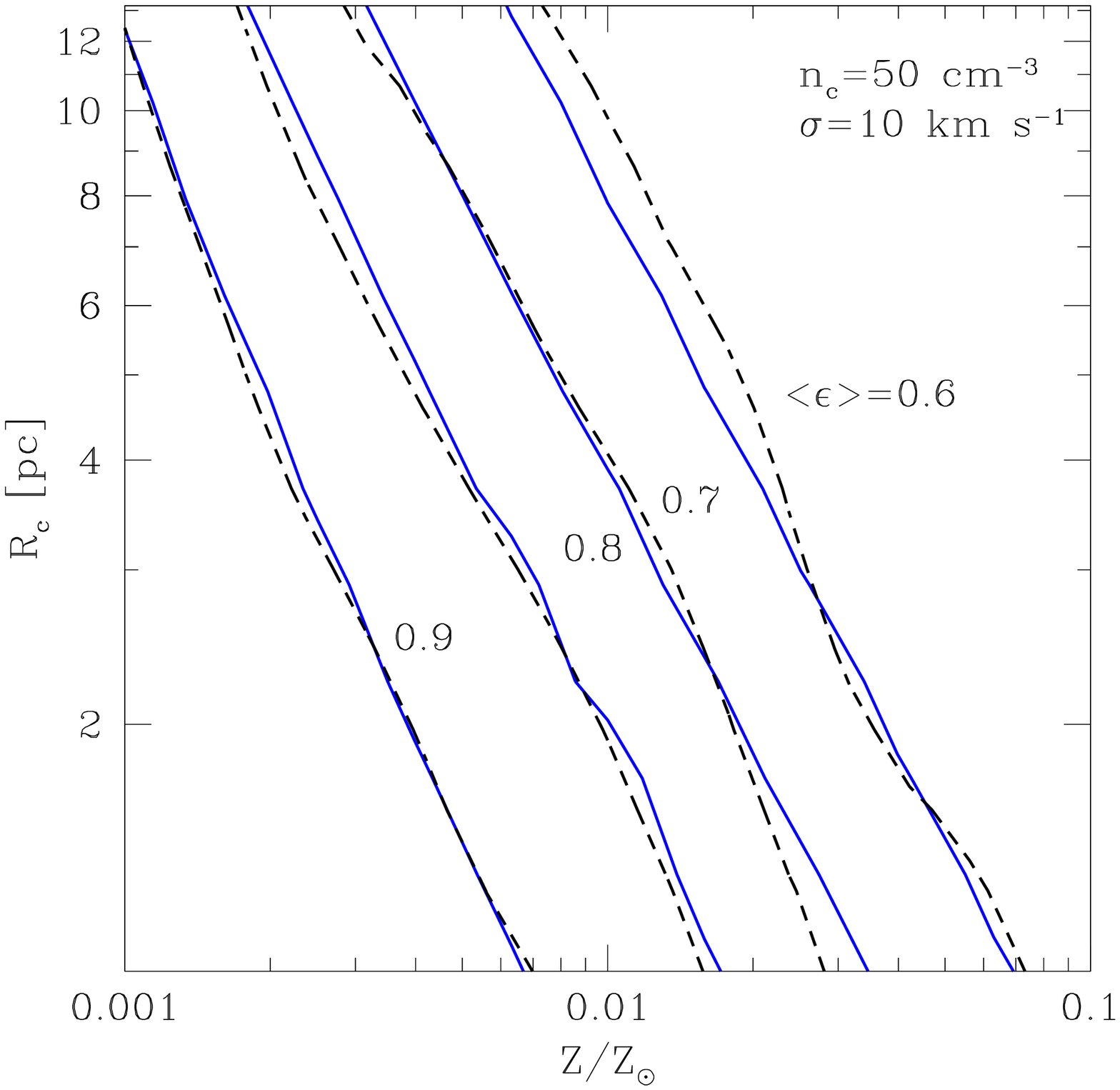,height=8cm}}
\caption{\label{fig:Z-R} Isocontours of
  $\beta$ (left) and mean elasticity $\langle \epsilon \rangle$
  (right) as a function of gas metallicity $Z$ and cloud size $R_c$.
  The solid lines refer to Monte Carlo simulation results; the dashed
  lines are the analytic solutions of eqs.~\ref{eq:med_el}. The cloud
  density is $n_c=50$ cm\mmm, the temperature $T=48$ K and $\sigma=10$
  km~s\m.}
\end{figure*}

We now use the more realistic prescription, eq.~(\ref{eq:rfm}) for the
elasticity of cloud-cloud collisions. The solid lines in
Fig.~\ref{fig:Z-R} show the isocontours of constant $\beta$ (left
panel) and mean elasticity $\langle \epsilon \rangle$ (right panel) as
a function of $Z$ and $R_c$ from the Monte Carlo simulations. The
cloud gas density is $n_c=50$ cm\mmm and the temperature $T=48$ K, the
typical values for a two-phase ISM with $Z=Z_\odot$ (see RFM). 

The value of $\beta$ is found to be related to elasticity similarly to
eq.~\ref{eq:main}, where now  $\langle \epsilon \rangle$ is the mean
elasticity.  The latter is defined by 

\begin{equation}
\left\{
\begin{array}{l}
\langle \epsilon \rangle = \int_0^\infty dv_r
p(\sigma,\beta,v_r)\epsilon(v_r,Z,n_c,R_c) \\[2mm]
\beta = 2  e^{(\langle\epsilon \rangle -1)}~~~~~~~~0.5 \le
\langle \epsilon \rangle\le 1  \label{eq:med_el}
\end{array}
\right.
\end{equation}
where $p(\sigma,\beta,v_r)$ is the probability distribution function 
for the modulus of the relative
velocity (rPDF). The values of $\beta$ and $\langle \epsilon \rangle$,
derived solving the system of eqs.~(\ref{eq:med_el}), are shown in
Fig.~\ref{fig:Z-R} (dashed lines).

The 1D rPDF is given by 
\begin{equation}
p_{1D}(v_r)=\int dv_1 \int dv_2 p_{1D}(v_1)p_{1D}(v_2) \delta[v_r-(v_1-v_2)]
\label{eq:p_vrel}
\end{equation}
Here, $\delta$ is the Dirac delta function and $p_{1D}(v) \propto
\exp[-(v/\sigma_{1D})^\beta]$ is the normalized 1D PDF.  It is easy to
solve eq.~(\ref{eq:p_vrel}) for the particular cases $\beta=2$ and
$\beta=1$. For $\beta=2$ the 1D rPDF is a Gaussian with
$\sigma_r=\sqrt{2} \sigma$. In the $\beta=1$ case, the rPDF is
\begin{equation}
p_{1D}(\sigma_{1D},\beta=1,v_r)={|v_r|+\sigma_{1D} \over 4 \sigma_{1D}^2} \exp\left({-{|v_r|
  \over \sigma_{1D}}}\right)
\label{eq:p_vrelb1}
\end{equation}
For arbitrary values of $\beta$ eq.~(\ref{eq:p_vrel}) has to be solved
numerically. Eq.~(\ref{eq:p_vrelb1}) shows that the functional form of
the 1D rPDF and therefore the rPDF $p(\sigma,\beta,v_r)$ changes
depending on $\beta$. Nevertheless, using Monte Carlo simulations we
find that the parametric function $p(\sigma,\beta,v_r) \propto v_r^2
\exp[-(v_r/\sigma_r)^{\beta_r}]$ with
$\sigma_r=\sigma(1+\beta^{-3})2^{1 \over \beta}$ and
$\beta_r=\beta(1+0.24\beta^{-3})$ is a good fit to rPDF. We have used
this parameterization to solve the system of eqs.~(\ref{eq:med_el}).

\begin{figure}
\centerline{\psfig{figure=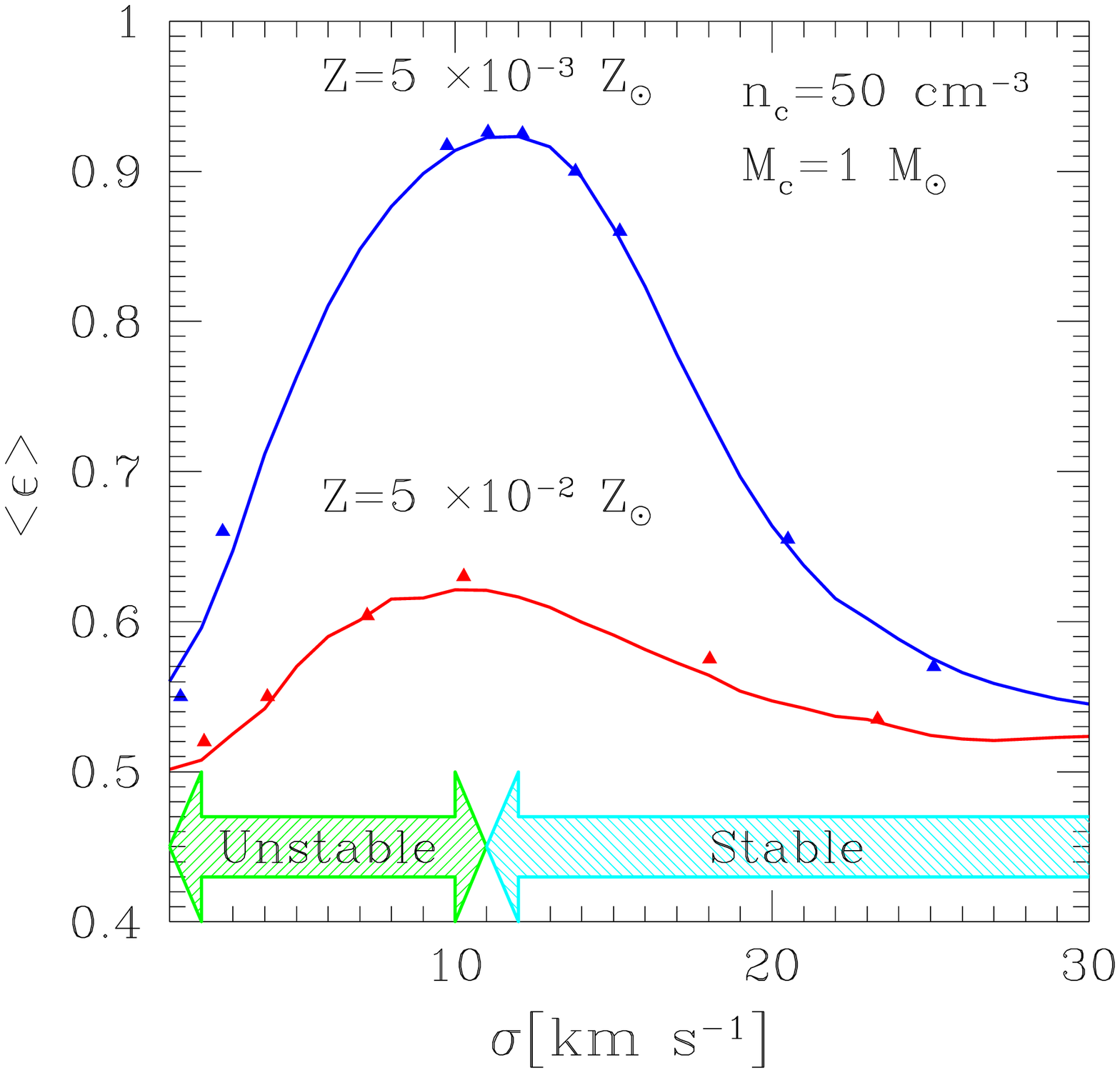,height=9cm}}
\caption{\label{fig:el-sig} Mean elasticity $\langle \epsilon
  \rangle$, as a function of cloud velocity dispersion $\sigma$. The
  solid lines show the analytical result obtained by solving
  eq.~(\ref{eq:med_el}) for $Z=5 \times 10^{-2} Z_\odot$ and $Z=5
  \times 10^{-3} Z_\odot$ (small integration errors are responsible
  for the wavy appearance of the lower line).  The triangles show the
  analogous relation as obtained from Monte Carlo simulations. It can
  be shown that if $\sigma<\sigma_c\simeq 11$ km~s\m, the PDF is
  unstable.}
\end{figure}

The mean elasticity, and therefore $\beta$, depend on $\sigma$. In
Fig.~\ref{fig:el-sig} we show $\langle \epsilon \rangle$, calculated
from eq.~(\ref{eq:med_el}), as a function of $\sigma$ for $Z=5 \times
10^{-2} Z_\odot$ and $Z=5 \times 10^{-3} Z_\odot$. The triangles show
the analogous relation obtained from Monte Carlo simulations. 

As discussed in \S~\ref{ssec:ccrev}, cloud-cloud collisions
are more elastic when their relative velocity is about 30 Km
s$^{-1}$. This can be understood if we recall that the elasticity of
the collision is proportional to the column density of the post-shock
radiative region, $\Sigma_{cool} \propto v_s t_{cool} \propto v_s^3
/\Lambda(T)$, where $v_s$ is the shock velocity, $t_{cool}$ the
cooling time of the shocked gas and $\Lambda(T)$ is the gas cooling
function. For small collision velocities the post-shock gas is cold
($T<10^4$ K) therefore the cooling time is relatively long. In this
regime the elasticity increases with the cloud relative velocity about
as $v_r^3$. When the cloud relative velocity is high enough that the
post-shock gas is heated above $T \sim 2 \times 10^4$ K, the cooling
time becomes very short because of the Ly$\alpha$ cooling. Therefore
the elasticity decreases steeply. The maximum of the mean elasticity
$\langle \epsilon \rangle$ at $\sigma \approx 11$ Km s$^{-1}$ is the
consequence of the dependence of the cloud-cloud collision elasticity
on the velocity of the clouds. In other words $\Sigma_{cool}$ is
maximum when $\sigma \approx 11$ Km s$^{-1}$.

\begin{figure}
\centerline{\psfig{figure=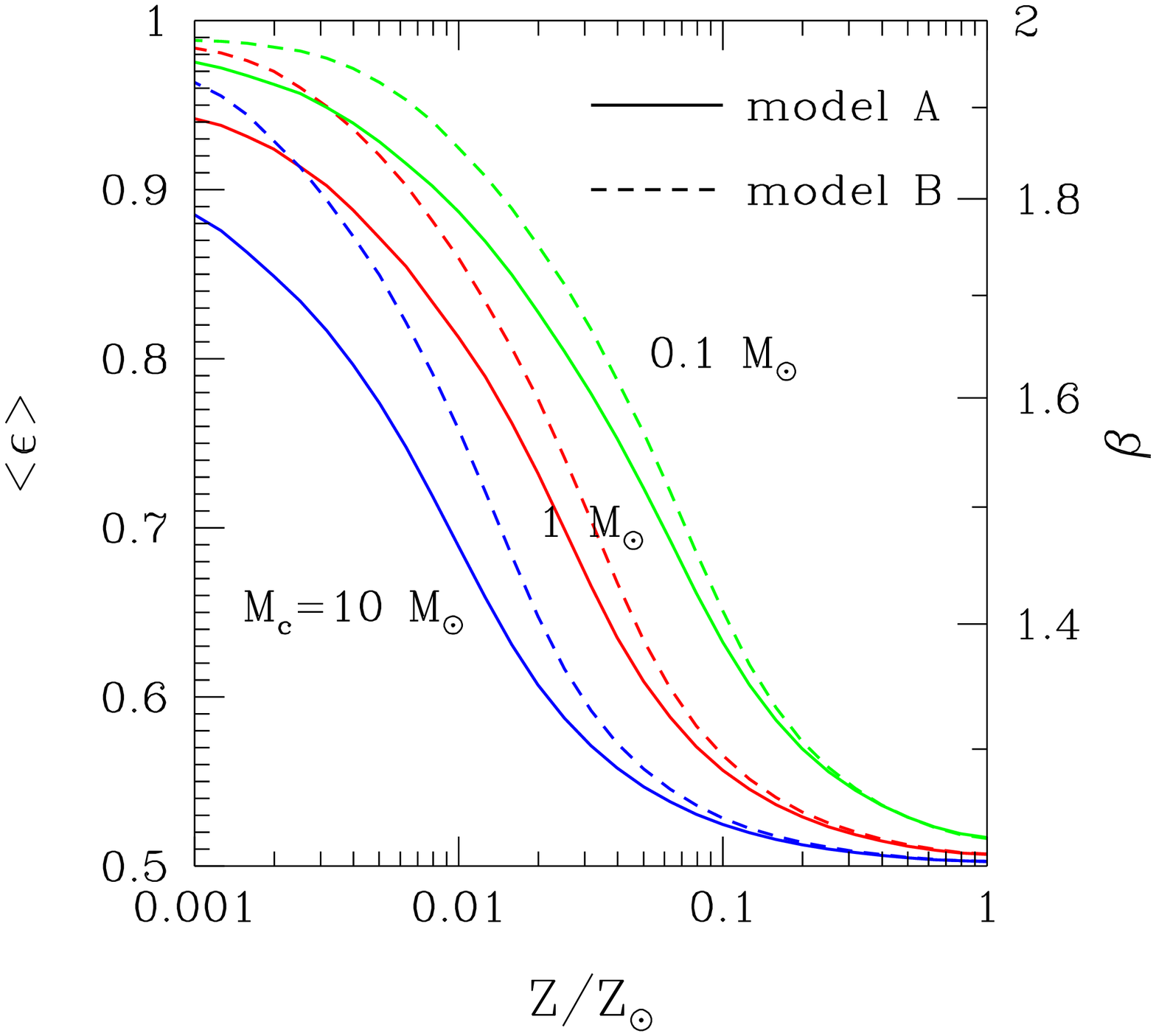,height=9cm}}
\caption{\label{fig:el-Z} Mean elasticity $\langle \epsilon \rangle$,
  as a function of the metallicity $Z$ (normalized to the solar
  metallicity), for $\sigma=11$ km~s\m. We assume that the clouds are
  in pressure equilibrium in a multiphase medium.  This assumption
  imply that the clouds have temperature $T=48$ K and density $n_c(Z)
  \simeq n_c(Z=1)\exp[{0.0834(Z^{-1/2}-1)}]$ (\ie, $P(Z)=n_cT \simeq
  P(Z=1)\exp[{0.0834(Z^{-1/2}-1)}]$, where $P(Z=1)=2400$ cm\mmm
  K). Here we assume $n_c(Z=1)=50$ cm\mmm. The solid lines show the
  mean elasticity assuming that cloud masses scale with metallicity as
  $M_c(Z)=M_c(Z=1)[n_c(Z)/n_c(Z=1)]^{-1/2}$ for $M_c(Z=1)=0.1,1,10$
  M$_\odot$ (Model A). The dashed lines show the analogous relation
  assuming $M_c(Z)=M_c(Z=1)[n_c(Z)/n_c(Z=1)]^{-2}$ for
  $M_c(Z=1)=0.1,1,10$ M$_\odot$ (Model B). The rationale for models A
  and B is explained in the text.}
\end{figure}

The dependence of $\langle \epsilon \rangle$ on $\sigma$ indicates
that if $\sigma<\sigma_c\simeq 11$ km~s\m the PDF cannot reach a
stationary state, \ie an instability occurs. Physically, this is due
to the fact that in this low-$\sigma$ regime an increase (decrease) of
$\sigma$ produces an increase (decrease) of $\langle \epsilon
\rangle$. As a result less (more) energy is dissipated and $\sigma$
increases (decreases) further. The simulation results shown for this
$\sigma$ range do not reflect equilibrium values but are given at a
time $t \sim 100$ Myr; eventually, depending on the parameters of the
simulation the PDF relaxes either to the stable high-$\sigma$ branch
or to $\sigma \sim 0$. The time scale for this process is essentially
given by the growth rate of the instability, which is particularly
fast for a low metallicity gas and vanishes as $Z$ is increased above
$Z \approx 0.1 Z_\odot$. The growth rate of the instability is
$\tau^{-1} \sim |E_i/\langle E \rangle - \langle \eta \rangle | \nu
\propto (1-\langle \epsilon \rangle) \nu$, where we remind that $\nu$
is the cloud collision rate. The triangles in the unstable region are
evolved from the same Maxwellian PDF with $\sigma_M=10$ km~s\m but
with different values for $E_i$. The fact that the triangles in the
unstable region lay very close to the solid line implies that
eq.~(\ref{eq:E}) is approximatively valid also when the system is not
in a steady state. If $E_i<E_i^{crit} \sim 121 [\Gamma(5/ \beta^*)/
\Gamma(3/\beta^*] \ln (2 / \beta^*)$ km$^2$ s\mm, where
$\beta^*=\beta(\sigma=11)$ is a function of $Z, M_c$ and $n_c$ (see
Fig.~\ref{fig:el-Z}), the PDF will be unstable.  Therefore, from
Fig.~\ref{fig:el-sig} we conclude that in a multiphase, low
metallicity ($Z \approx 5 \times 10^{-3} Z_\odot$) ISM the PDF should
be close to a Maxwellian ($\beta = 2$) with velocity dispersion
$\sigma \simgt 11$ km~s\m; in more metal rich systems ($Z \simgt 5
\times 10^{-2} Z_\odot$) instead we expect to observe an almost
exponential tail of the PDF with $\beta \approx 1.2$. If $Z \ga 0.1$,
the instability of the PDF is not fast enough to put strong
constraints on the value of $\sigma$.  We will discuss in
\S~\ref{sec:obs} the observational implications of these results.

Finally, we estimate the typical cloud mass as a function of the
metallicity.  As discussed in RFM, a multiphase ISM in thermal
equilibrium can exist in a narrow range of thermal pressure. The
pressure range depends on the gas metallicity but the temperature of
the clouds (CNM) is $T \sim 48$ K, independent of the metallicity.
This implies that exist a relationship between the cloud density $n_c$
and $Z$.  If we assume a cloud temperature $T=48$ K and pressure
equilibrium of the multiphase ISM we find
\begin{equation}
n_c(Z) \simeq n_c(Z=1)\exp[{0.0834(Z^{-1/2}-1)}].
\label{eq:peq}
\end{equation}
Adopting  this relation we explore two different models: 

\underline{\it Model A} Since diffuse clouds are not self gravitating, 
we assume that their mass is a small (constant) fraction of the Jeans mass
$M_J=(110 M_\odot) T^{3/2} n_c(Z)^{-1/2}.$
The temperature of the cloud is approximately constant
($T=48$ K), therefore the mass is $M_c \propto n_s(Z)^{-1/2}$
and the radius $R_c \propto n_c(Z)^{-1/2}$.

\underline{\it Model B} Alternatively, we can derive the cloud size by
imposing that the H$_2$ abundance in the cloud is negligible. If the
cloud is optically thin to the H$_2$ photodissociating radiation in
the Lyman-Werner bands (\ie, has a column density $\Sigma_c=n(Z)R_c \simlt (5
\times 10^{14}~{\rm cm}^{-2})$), the H$_2$ abundance will be $x_{H_2}
\simlt 10^{-9}$ if the UV background in the galaxy has values
comparable to the Milky-Way. We therefore assume that the column
density of the clouds is a (constant) fraction of $5 \times
10^{14}~{\rm cm}^{-2}$. In this case we find the scaling laws: $R_c \propto
n_c(Z)^{-1}$ and $M_c \propto n_c(Z)^{-2}$.

In Fig.~\ref{fig:el-Z} we show the maximum (at $\sigma=11$ km
s$^{-1}$) mean elasticity $\langle \epsilon \rangle$ as a function of
the gas metallicity, $Z$. The three solid lines show the mean
elasticity assuming that cloud masses scale with metallicity as
$M_c(Z)=M_c(Z=1)[n_c(Z)/n_c(Z=1)]^{-1/2}$ for $M_c(Z=1)=0.1,1,10$
M$_\odot$ (\ie, Model A). The dashed lines show the analogous relation
assuming $M_c(Z)=M_c(Z=1)[n_c(Z)/n_c(Z=1)]^{-2}$ for
$M_c(Z=1)=0.1,1,10$ M$_\odot$ (\ie, Model B). The mean elasticity
increases when the gas metallicity decreases as a consequence of the
lower cooling rate of metal-deficient gas. When the mass of the clouds
is small the mean elasticity is larger because of the smaller cloud
column density, $\Sigma_c$ (we have seen in \S~\ref{ssec:ccrev} that
the cloud-cloud collision elasticity is inversely proportional to
$\Sigma_c$). The difference between models A and B is also caused
by the dependence of the cloud column density on the gas
metallicity. In Model A $\Sigma_c(Z) \propto n_c(Z)^{1/2}$ and in
Model B $\Sigma_c(Z)$ is constant.


\section{Observational Links}\label{sec:obs}

The results presented above allow to make a number of predictions
concerning the ISM velocity PDF which can be tested against
observations.  The first attempts to determine the PDF of the ISM date
back to the early 50s. These studies used optical absorption lines
toward OB stars or HI 21 cm emission or absorption line to measure the
velocity PDF of local gas at distances $d \la 1$ kpc.  More recently
the question of the velocity PDF has received much less attention.  In
the following, we will briefly summarize the available observational
results; we particularly emphasize the point that the large majority
of them indicate that a exponential PDF at large $\sigma$ is a much
better fit than a Gaussian one to the velocity data.
 
Adams (1949) compiled a catalog including 230 Ca K absorption line
velocities produced by ``clouds'' along the line of sight to nearby OB
stars. He found 150 components towards 120 stars with $d < 500$ pc and
80 components towards 43 stars with $d > 500$ pc.  Using such catalog,
Blaauw (1952)\nocite{Blaauw:52} and Takakubo
(1967)\nocite{Takakubo:67}, showed that a single exponential could fit
the velocity PDF of local interstellar clouds better than a single
Gaussian.  Analyzing the same data, Huang (1950)\nocite{Huang:50} and
Kaplan (1954)\nocite{Kaplan:54} preferred a fit with a $v^{-1}$
function; Siluk \& Silk (1974)\nocite{Siluk:74} fitted the
high-velocity tail of the PDF with a $v^{-3}$ function.  Another set
of optical interstellar lines, toward 132 stars was observed by Munch
(1957)\nocite{Munch:57}.  He used a different method, the ``doublet
ratio method'' (Ca H and K or NaD2/D1). His conclusions confirm
Blaauw's results.

Mast \& Goldstein(1970)\nocite{Mast:70} observed the radial velocities
of a sample of 268 HI 21 cm emission clouds. In order to clearly
resolve the clouds in velocity, they chose a sample of high latitude
line of sights and estimate the peculiar velocity distribution after
correction for solar motion and differential galactic rotation. It is
evident from their Fig.~4 that an exponential fits better the data
than a Gaussian.  Falgarone and Lequeux (1973)\nocite{Falgarone:73},
using data from Radhakrishnan \etal (1972
a,b,c,d)\nocite{Radhakrishnan:72}, Goss \& Radhakrishnan (1972) and
Hughes, Thompson \& Colvin (1971)\nocite{Hughes:71} analyzed the PDF
of the ISM in the solar neighborhood. Their results are in agreement
with those of Mast \& Goldstein(1970) but their sample is too limited
to discriminate the detailed shape of the distribution.  Crovisier
(1978)\nocite{Crovisier:78} studied the kinematics of nearby HI clouds
using the Nancay~21-cm absorption survey (Crovisier \etal
1978\nocite{Crovisieretal:78}). The sample consisted of about 300
cloud radial velocities. They also confirmed that the 1D PDF is close
to exponential, although their conclusion suffers of possible
contamination from a population of intermediate-velocity clouds.
Dickey, Terzian, \& Salpeter (1978)\nocite{Dickey:78} obtained the
radial velocity distribution of clouds from the Arecibo 21-cm
emission/absorption survey.  Unfortunately, the statistics is too
small to clearly discriminate the functional form of the PDF.  Kim
\etal (1998)\nocite{Kim:98} measured the PDF for motions out of the
plane of the Large Magellanic Cloud. The peculiar velocity PDF, on
scales larger than 15 pc, is well fitted by an exponential.

In a slightly different context, Miesch, Scalo \& Bally
(1999)\nocite{Miesch:99} estimated the PDF of molecular line centroid
velocity fluctuations for several nearby molecular clouds with active
internal star formation.  The data consist of over 75,000 $^{13}$CO
line profiles divided among 12 spatially and/or kinematically distinct
regions.  These regions range in size from less than 1 to more than 40
pc and show substantially supersonic motions. They find that 3 regions
(all in Mon R2) exhibit nearly Gaussian centroid PDFs, but the other 9
show nearly exponential PDFs.

In spite of this impressive amount of evidences, many authors have
continued to make the assumption when modeling or interpreting ISM
data that the velocity PDF for the diffuse and molecular clouds is
Gaussian. This bias is probably due to the expectations from
incompressible turbulence models of the ISM, which predict that the
PDF is nearly Gaussian.  For systems whose metallicity is larger
than about 1/20 of solar, our model predicts instead that the velocity
PDF should approximate an exponential distribution.  The physical
reason for this discrepancy resides in the dissipative nature of cloud
collisions.  From our results, an exponential PDF can only be obtained
if the ensemble averaged elasticity is 0.5, indicating perfectly
inelastic collisions.  Hence, kinetic energy stored in turbulent
motions is efficiently converted in thermal energy and radiated away
in shocks arising in colliding flows. The (supersonic) turbulence
decays very rapidly and a corresponding high rate of energy deposition
is required to stir the gas. This is in agreement with the findings of
up-to-date numerical simulations in the context of star formation
(e.g. Smith, MacLow \& Heitsch 2000; Padoan \& Nordlund 1999).

Finally, we would like to comment on the dynamical instability
discussed above (see Fig.~\ref{fig:el-sig}). If values of $\sigma$
corresponding to the unstable $\sigma < 11$ km~s\m region are deduced
from observations of a low metallicity system, because of the runaway
character of the instability, $\sigma$ should rapidly decrease or
increase assuming a constant SN energy injection rate. If $\sigma$
decreases, leading to a loss of pressure support and hence to overall
contraction, is likely to drive a similarly rapid increase of the star
formation rate and energy injection from SN explosions. This will
cause $\sigma$ to start increasing. An increasing $\sigma$ will lead
to overall expansion of the ISM and the star formation rate and the SN
energy injection will decrease back to the the starting value.
Therefore $\sigma$ starts decreasing again unless the energy injection
rate is sufficient to maintain $\sigma > 11$ km~s\m.  As a result a
``bursty'' star formation mode will occur, unless $\sigma$ is
increased up to the stable region, thus stabilizing the star formation
rate on a higher level. The period of the burst cycle should increase
with the gas metallicity because the timescale of the instability
does.  A ``bursty'' mode of star formation has indeed been observed in
some low surface brightness (LSB) dwarf galaxies (Schombert, McGaugh
\& Eder 2001).  Thus, the dynamical instability occurring together
with the dissipation of kinetic energy investigated here might provide
an interesting self-regulation of the star formation activity in
low-metallicity and low surface brightness galaxies. The suggested
self-regulation mechanism is rather speculative given the simplicity
of our model and should be tested using more realistic numerical
simulations.

\section{Discussion}\label{sec:disc}

Turbulent supersonic flows and gas self-gravitation create structures
in the ISM that are very complex and hierarchically structured. It is
still under debate whether interstellar clouds are only transient
phenomena produced by dynamical fluctuations or can be represented as
long-living clouds.  Numerical simulations (Wada,
2001)\nocite{Wada:01} and observations (Schwarz \& van Woerden,
1974)\nocite{Schwarz:74} show that the morphology of the atomic
component of the ISM is a complex network of filamentary structures.
The filaments are probably produced by oblique shock collisions and
shear from differential galactic rotation. The typical size of the
filaments in the simulations is about 1 pc, consistent with the
observed typical size of atomic interstellar clouds. Contrary to
self-gravitating molecular clouds, the atomic clouds do not show
evident substructures.

In our simple toy model clouds do not coalesce or are destroyed, and
their mass, size and number density remain constant. Our assumptions
are not physically justified but are the correct choice for a
statistical description of the ISM. 

The outcome of slow inelastic collisions could produce cloud
coalescence. If this process is important, eventually the cloud mass
reaches the Jeans mass and the atomic cloud becomes a self-gravitating
molecular cloud. Perhaps this process, in conjunction with
fragmentation of dense gas shells produces by SN explosions, is
responsible for the regulation of the gas ratio between the atomic and
molecular phases of the ISM. Simulations show that small clouds are
probably destroyed if they are passed by supersonic shocks. In this
process mass is exchanged between atomic clouds and the intercloud
diffuse gas. Indeed from numerical simulations it is evident that
clouds are not long lived entities but instead they are continuously
formed and destroyed.  But despite the complex and time dependent
morphology of the ISM, statistical analysis of the cloud properties in
numerical simulations shows that the typical cloud size, cloud mass,
and cloud number density have constant equilibrium values (Wada \&
Norman, 2001)\nocite{WadaN:01}.  Therefore our model is not physically
realistic but our simple assumptions are justified in a statistical
sense. For this reason we cannot study the ISM morphology or the
physics that regulates mass exchange between different phases of the
ISM. But we believe that our model is reliable for a statistical study
of kinetic energy dissipation in ``cloud'' collisions. At least on
scales where the isotropy assumption in the ISM holds.

Because of the aforementioned arguments, we suspect that adding
complexity to the model (for instance assuming non homogeneous density
clouds or a cloud mass spectrum) would not help substantially in
making the results more realistic and reliable.  Nevertheless, in a work
currently in preparation, we are using 2D high-resolution hydrodynamic
simulations to model the ISM in galaxies with primordial (metal-poor)
composition.  These simulations should be able to test the quality of
the results presented in this paper.

Finally we discuss to which extent the results of this work can be
extended to different systems, namely sub-clumps in molecular clouds
and when the main turbulent energy source is different from SN
explosions.  

Observations of the velocity PDF in molecular cloud cores (Miesch,
Scalo \& Bally, 1999)\nocite{Miesch:99} have shown that the 1D PDF is
better represented by an exponential rather than Gaussian
distribution. Based on the results of our model, that applies to
atomic clouds, we have suggested that kinetic energy dissipation in
collision of sub-clumps is at the origin of the exponential velocity
distribution. Even if we believe that an analogy between atomic clouds
in the ISM and sub clumps in molecular clouds is possible, we have to
keep in mind important differences.  Molecular clouds, are self
gravitating and harbor smaller clumps for which size-mass-velocity
relations have been derived from observations (\eg, Larson,
1981)\nocite{Larson:81}. The inter-clump gas can be atomic especially
in the outermost regions. Our assumption of homogeneity and
isothermality for clouds of same sizes and the derived scalings can
serve as a first-order approach to the real hierarchically structured
molecular cloud. More important, in our model we do not include
molecular chemistry and cooling. Note that, if molecular hydrogen is
not dissociated, clump collisions are inelastic (because of radiative
cooling from collisionally excited H$_2$ rotational and vibrational
levels) even if the gas has primordial composition (metal-free).

One of the results discussed in \S~\ref{sec:res} is that the shape of
the PDF does not depend on the kinetic energy input prescription if
the energy input event (SN explosion in our case) is much larger than
the energy dissipated by a typical collision event. This means that
our results should remain valid even if the main kinetic energy source
that fuels the ISM turbulence is not SN explosions. Stellar winds are
are important energy sources. Chimney and superbubbles created by
multiple SN explosions in a stellar cluster can reduce the amount of
energy input in the ISM and produce, instead, gas outflows from the
ISM to the intergalactic medium.  Finally Wada \& Norman
(1999)\nocite{WadaN:99} have shown that the rotational energy of the
disk is converted into turbulent energy of the ISM because of the
combined effects of gas self-gravity and disk differential rotation.

\section{Summary}\label{sec:sum}

We have investigated the kinetic energy of the ISM in galaxies as a
function of their gas metallicity. We use a simple closed model where
the energy injection from SN explosions balance the energy dissipated
in cloud collision. Using Monte Carlo simulations, coupled with a
simple (but statistically motivated) model for the clouds and previous
numerical results on the elasticity of collisions derived by RFM, we
have been able to link the properties of the velocity PDF with the ISM
kinetic energy dissipation.  In particular, we find that:

$\bullet$ The slope, $\beta$, of the velocity PDF only depends on the
value of the ensemble averaged elasticity $\langle \epsilon \rangle$
as $\beta = 2 \exp(\langle\epsilon \rangle -1)$.

$\bullet$ The knowledge of $\beta$ and of the velocity dispersion
$\sigma$ of the gas allows to determine the specific kinetic energy
input from SNe, $E_i$ through the simple relation $E_i \approx 1.58
\sigma^2 \ln (2 / \beta)/(\beta-0.947)$; in steady state, this input
is equal to the energy dissipated by collisions.

$\bullet$ We predict that (see Fig.~\ref{fig:el-sig}) in a multiphase, low
metallicity ($Z \approx 5 \times 10^{-3} Z_\odot$) ISM the PDF should
be close to a Maxwellian ($\beta = 2$) with velocity dispersion
$\sigma \simgt 11$ km~s\m; in more metal rich systems ($Z \simgt 5
\times 10^{-2} Z_\odot$), instead we expect to observe exponential
PDFs with $\beta \approx 1.2$.

$\bullet$ We have pointed out that, contrary to what is usually
assumed, the available data firmly allow to conclude that the Galactic
ISM velocity PDF is not Gaussian but has rather an exponential
shape. This is in perfect agreement with our predictions.

$\bullet$ Low metallicity systems with $\sigma < 11$ km~s\m are
dynamically unstable on time scales $\tau \propto [(1-\langle \epsilon
\rangle)\nu]^{-1}$ where $\nu$ is the cloud collision rate. This
systems are probably undergoing a "bursty" mode of star formation.
Indeed, some observations suggest that LSB dwarf galaxies show signs
of repeated and weak bursts of star formation which shape their
irregular morphology appearance (Schombert, McGaugh \& Eder 2001).  On
the contrary, a quiescent star formation mode as inferred for the
Milky Way, should be associated with an ISM velocity dispersion
$\sigma \approx 11$ km~s\m and a relatively high star formation rate.


\label{lastpage}
\end{document}